\def\rfr#1{eq. (\ref{#1})}

\def\virg#1{``#1''}

\def\eqi{\begin{equation}}
\def\eqf{\end{equation}}
\def\eqia{\begin{eqnarray}}
\def\eqfa{\end{eqnarray}}
\def\rp#1#2{{#1\over#2}} \def\lb#1{\label{#1}}

\documentclass[11pt]{article}%{ws-ijmpd}
\usepackage{amsmath,amsthm,amscd,amssymb}
\usepackage{latexsym}
\usepackage{graphicx,epsfig}

\begin{document}

\noindent{\bf \LARGE{On possible \textit{a-priori} \virg{imprinting} of General Relativity itself on the performed Lense-Thirring tests with LAGEOS satellites}}
\\
\\
\\
{L. Iorio$^{\ast}$\\
{\it $^{\ast}$INFN-Sezione di Pisa. Address for correspondence: Viale Unit$\grave{a}$ di Italia 68
70125 Bari (BA), Italy.  \\ e-mail: lorenzo.iorio@libero.it}

\vspace{4mm}

\begin{abstract}
The impact of possible a-priori \virg{imprinting} effects of general relativity itself on recent attempts to measure the Lense-Thirring precessions with the LAGEOS satellites orbiting the Earth and the terrestrial geopotential models by the dedicated mission GRACE is investigated. It is analytically shown that general relativity, not explicitly solved for in the GRACE-based models, may \virg{imprint} their even zonal harmonic coefficients $J_{\ell}$ at a non-negligible level, given the present-day accuracy in recovering them. This translates into a bias of the LAGEOS-based relativistic tests  as large as the Lense-Thirring effect itself. Further analyses should include general relativity itself in the GRACE data processing by explicitly solving for it.
\end{abstract}

\section{Introduction}
The term \virg{gravitomagnetism} \cite{Thorne,Rin,Mash} (GM) denotes those gravitational phenomena concerning orbiting test particles, precessing gyroscopes, moving clocks and
atoms and propagating electromagnetic waves \cite{Tar,Sciaf} which, in the framework of the Einstein's
General Theory of Relativity  (GTR), arise from non-static distributions of matter and energy. In the weak-field and slow motion approximation, the Einstein field equations  of GTR, which is a highly non-linear Lorentz-covariant tensor theory of gravitation, get
linearized \cite{line}, thus looking like  the Maxwellian equations of electromagntism. As a consequence, a \virg{gravitomagnetic}
field $\vec{B}_g$, induced by the off-diagonal components $g_{0i},\ i = 1,2,3$ of the space-time metric
tensor related to mass-energy currents, arises. In particular, far from a localized rotating body with angular
momentum $\vec{S}$ the gravitomagnetic field can be written as \cite{PLA}
\eqi \vec{B}_g(\vec{r}) = \rp{G}{cr^3}\left[\vec{S} -3\left(\vec{S}\cdot\hat{r}\right)\hat{r} \right],\lb{gmfield}\eqf where $G$ is the Newtonian gravitational constant and $c$ is the speed of light in vacuum. It affects, e.g., a test particle moving with velocity $\vec{v}$ with a non-central acceleration \cite{PLA}
\eqi \vec{A}_{\rm GM}=\left(\rp{\vec{v}}{c}\right)\times \vec{B}_g.\eqf It is the cause of the so-called Lense-Thirring\footnote{According to a recent historical analysis, it should be more correct to speak about an Einstein-Thirring-Lense effect \cite{Pfi}.} effect \cite{LT}, which is one of the most famous and empirically investigated GM features; another one is the gyroscope precession \cite{Pu,Sci}, goal of the Gravity Probe B (GP-B) mission \cite{gpb} whose data analysis is still ongoing \cite{ANA}.

The Lense-Thirring effect consists of small secular precessions of the longitude of the ascending node $\Omega$ and the argument of pericenter $\omega$ of the orbit of a test particle in geodesic motion around a slowly rotating body with angular momentum $\vec{S}$; they are
\eqi\dot\Omega_{\rm LT}=\rp{2GS}{c^2 a^3(1-e^2)^{3/2}},\ \dot\omega_{\rm LT}=-\rp{6GS\cos I}{c^2 a^3(1-e^2)^{3/2}},\eqf where $a$ is the semimajor axis of the satellite's orbit, $e$ is its eccentricity and $I$ is the inclination of the orbital plane to the equatorial plane of the central body.

Concerning the possibilities of measuring it in the terrestrial gravitational field, soon after the dawn of the space age with the launch of Sputnik in 1957 it was proposed by Soviet scientists to directly test the Lense-Thirring effect with artificial satellites orbiting the Earth. In particular, V.L. Ginzburg \cite{Ginz0,Ginz1,Ginz2} proposed to use the perigee of a terrestrial spacecraft in highly elliptic orbit, while  A.F. Bogorodskii \cite{Bogo} considered also the node.  In 1977-1978 Cugusi and Proverbio \cite{Cugu1,Cugu2} suggested to use the passive geodetic satellite LAGEOS, in orbit around the Earth since 1976 and tracked with the Satellite Laser Ranging (SLR) technique, along with the other existing laser-ranged targets to measure the Lense-Thirring node precession.
Since such earlier studies it was known that a major source of systematic error is represented by the fact that the even ($\ell=2,4,6,...$) zonal ($m=0$) harmonic coefficients $J_{\ell}, \ell=2,4,6$ of the multipolar expansion of the classical part of the terrestrial gravitational potential, accounting for its departures from spherical symmetry  due to the Earth's diurnal rotation, induce competing secular precessions  of the node and the perigee of satellites \cite{Kau} whose nominal sizes are several orders of magnitude larger than the Lense-Thirring ones.
In the case of the node, the largest precession is due to the first even zonal harmonic $J_2$
\eqi \dot\Omega_{J_2}=-\rp{3}{2}n\left(\rp{R_{\oplus}}{a}\right)^2\rp{\cos I J_2}{(1-e^2)^2},\eqf where $R_{\oplus}$ is the Earth's mean equatorial radius and $n\doteq\sqrt{GM_{\oplus}/a^3}$ is the satellite's Keplerian mean motion.
For the other higher degrees the analytical expressions are more involved; since they have already been published in, e.g., Ref.~\cite{Ior03}, we will not show them here.

Tests have started to be effectively performed about 15 years ago by Ciufolini and coworkers \cite{Ciuetal96} with  the
LAGEOS and LAGEOS II satellites\footnote{LAGEOS II was launched in 1992.}, according to a strategy by
Ciufolini \cite{Ciu96} involving the use of a suitable linear combination of the nodes $\Omega$ of both
satellites and the perigee $\omega$ of LAGEOS II in order to remove the impact of the first two multipoles of the non-spherical gravitational potential of the Earth. Latest tests have been reported by Ciufolini and Pavlis \cite{Ciu04,CiuPeron}, Lucchesi \cite{Luc07} and  Ries and coworkers \cite{Riesetal} with only the nodes of both the satellites according to a combination of them explicitly proposed by Iorio\footnote{See also Refs. \cite{Pav02,Ries03a,Ries03b}.} \cite{Iornodi}. The total uncertainty reached is still matter of debate \cite{crit1,crit2,crit3,crit4,crit5,IorioSSRV,IorioCEJP} because of the lingering uncertainties in the Earth's multipoles and in how to evaluate their biasing impact; it may be as large as $\approx 20-30\%$ according to conservative evaluations \cite{crit1,crit4,crit5,IorioSSRV,IorioCEJP}, while more optimistic views \cite{Ciu04,CiuPeron,Riesetal} point towards $10-15\%$.

To be more specific, the node-only combination used in the latest tests is
\eqi \dot\Omega^{\rm LAGEOS} + c_1\dot\Omega^{\rm LAGEOS\ II},\ c_1=0.544.\lb{combi}\eqf It was designed to remove the effects of the static and time-varying components of $J_2$, so that \rfr{combi} is affected by the remaining even zonals of higher degree $J_4, J_6,...$.
   The gravitomagnetic trend given by \rfr{combi} amounts to 47.8 milliarcseconds year$^{-1}$ (mas yr$^{-1}$ in the following) since the Lense-Thirring node precessions for the LAGEOS satellites are 30.7 mas yr$^{-1}$ (LAGEOS) and 31.5 mas yr$^{-1}$ (LAGEOS II).
   The Lense-Thirring signal is usually extracted from long time series of computed\footnote{Actually, the nodes are not directly measurable quantities, so that speaking of \virg{residuals} is somewhat improper.} \virg{residuals} of the nodes of  LAGEOS and LAGEOS II obtained by processing their data with a suite of dynamical force models which purposely do not encompass  the gravitomagnetic force itself \cite{LucBal06,Luccc07}. The action of the even zonals is accounted for by using global solutions for the Earth's gravity field, in which general relativity has never been explicitly solved for\footnote{For a critical discussion of such an issue, see Ref.~\cite{Nor01}.}, produced by several institutions around the world from data of dedicated satellite-based missions like GRACE\footnote{See on the WEB http://icgem.gfz-potsdam.de/ICGEM/ICGEM.html.} \cite{GRACE}.

GRACE recovers the spherical harmonic coefficients of the geopotential  from both the tracking of the two satellites
by GPS  and  the observed intersatellite
distance variations \cite{eigengrace02s}. The possible \virg{memory} effect of the gravitomagnetic force in the satellite-to-satellite tracking was preliminarily addressed
in Ref.~\cite{crit1}. Here we will focus on the  \virg{imprint} coming from the GRACE orbits which is more important for us because it mainly resides in the low degree even zonals.
\section{A-priori \virg{imprinting} of General Relativity on the GRACE-based models}
Concerning that issue, Ciufolini and Pavlis   write in Ref.~\cite{crit2} that such a kind of leakage of the Lense-Thirring signal itself into the even zonals retrieved by GRACE is completely negligible because the GRACE satellites move along (almost)  polar orbits. Indeed, for perfectly polar ($I=90$ deg) trajectories, the gravitomagnetic force is entirely out-of-plane, while the perturbing action of the even zonals is confined to the orbital plane itself. According to Ciufolini and Pavlis \cite{crit2}, the deviations of the orbit of GRACE from the ideal  polar  orbital configuration would have negligible consequences on the \virg{imprint} issue. In particular, they write: \virg{the values of the even zonal harmonics
determined by the GRACE orbital perturbations are substantially independent
on the a priori value of the Lense–Thirring effect. [...] The small deviation from a polar orbit of the GRACE satellite, that
is $1.7\times 10^{-2}$ rad, gives only rise, $at\ most$, to a very small correlation with
a factor $1.7\times 10^{-2}$}.
The meaning of such a statement is unclear; anyway, we will show below that such a conclusion is incorrect.

The relevant orbital parameters of GRACE are quoted in Table \ref{tibulo};
\begin{table}
\caption{Orbital parameters of GRACE and its Lense-Thirring node precession. Variations of the orders of about 10 km in the semimajor axis $a$ and 0.001 deg in the inclination $I$ may occur, but it turns out that they are irrelevant in our discussion. (http://www.csr.utexas.edu/grace/ground/globe.html). \label{tibulo}
}
\centering
\bigskip
\begin{tabular}{@{}llll@{}}
\noalign{\smallskip}\hline\noalign{\smallskip}
$a$ (km)& $e$ & $I$ deg & $\dot\Omega_{\rm LT}$ (mas yr$^{-1}$)\\
 \noalign{\smallskip}\hline\noalign{\smallskip}
6835 & 0.001 & 89.02  & 177.4 \\
 \noalign{\smallskip}\hline\noalign{\smallskip}

\end{tabular}
\end{table}
the orbital plane  of GRACE is, in fact, shifted by 0.98 deg from the ideal polar configuration, and, contrary to what claimed in Ref.~\cite{crit2}, this does matter because its classical secular node precessions are far from being negligible with respect to our issue.
The impact of the Earth's gravitomagnetic force on the even zonals retrieved by GRACE can be quantitatively evaluated by computing the \virg{effective} value\footnote{It must be recalled that  $J_{\ell}=-\sqrt{2\ell +1}\ \overline{C}_{\ell 0}$, where $\overline{C}_{\ell 0}$ are the normalized gravity coefficients.}  $\overline{C}^{\rm LT}_{\ell 0}$ of the normalized even zonal gravity coefficients which would induce classical secular node precessions for GRACE as large as those due to its Lense-Thirring effect, which is independent of the inclination $I$.
To be more precise, $\overline{C}^{\rm LT}_{\ell 0}$ come from solving the following equation which connects the classical even zonal precession of degree $\ell$ $\dot\Omega_{J_{\ell}}\equiv \dot\Omega_{.\ell}J_{\ell}$ to the Lense-Thirring node precession $\dot\Omega_{\rm LT}$
\eqi \dot\Omega_{.\ell}J_{\ell}=\dot\Omega_{\rm LT}.\eqf In it \eqi \dot\Omega_{.\ell}=f(a,e,I; R_{\oplus},GM_{\oplus})\eqf are the coefficients of the classical node precessions depending on the satellite's orbital parameters and on the Earth's radius and mass.
Table \ref{tibulo2}
\begin{table}
\caption{ Effective \virg{gravitomagnetic}  normalized gravity coefficients for GRACE ($\ell=4,6;\ m=0$). They have been obtained by comparing the GRACE classical node precessions to the Lense-Thirring rate. Thus, they may be viewed as a quantitative measure of the leakage of the Lense-Thirring effect itself into the second and third  even zonal harmonics of the global gravity solutions from GRACE. Compare them with the much smaller calibrated errors in $\overline{C}_{40}$ and $\overline{C}_{60}$ of the GGM03S model \cite{ggm03} of Table \ref{tibulo3}. \label{tibulo2}
}
\centering
\bigskip
\begin{tabular}{@{}ll@{}}
\noalign{\smallskip}\hline\noalign{\smallskip}
$\overline{C}_{40}^{\rm LT}$ & $\overline{C}_{60}^{\rm LT}$ \\
 \noalign{\smallskip}\hline\noalign{\smallskip}
$2.23\times 10^{-10}$ & $-2.3\times 10^{-11}$ \\
 \noalign{\smallskip}\hline\noalign{\smallskip}

\end{tabular}
\end{table}
lists $\overline{C}^{\rm LT}_{\ell 0}$ for  degrees $\ell=4,6$, which are the most effective in affecting the combination of \rfr{combi}.
Thus, the gravitomagnetic field of the Earth contributes to the value of the second even zonal of the geopotential retrieved from the orbital motions of GRACE by an amount of the order of $2\times 10^{-10}$, while for $\ell=6$ the imprint is one order of magnitude smaller. Given the present-day level of accuracy of the latest GRACE-based solutions, which is of the order of $10^{-12}$ (Table \ref{tibulo3}), effects as large as those of Table \ref{tibulo2} cannot be neglected. Thus, we conclude that the influence of the Earth's gravitomagnetic field on the low-degree even zonal harmonics of the global gravity solutions  from GRACE may exist, falling well within the present-day level of measurability.
\begin{table}
\caption{Calibrated errors in the solved-for normalized gravity coefficients $\overline{C}_{40}$ and $\overline{C}_{60}$ according to the GGM03S global gravity solution by CSR \cite{ggm03}. They can be publicly retrieved at http://icgem.gfz-potsdam.de/ICGEM/ICGEM.html. Compare them with the much larger \virg{gravitomagnetic} imprinted coefficients of Table \ref{tibulo2}. \label{tibulo3}
}
\centering
\bigskip
\begin{tabular}{@{}ll@{}}
\noalign{\smallskip}\hline\noalign{\smallskip}
$\sigma_{\overline{C}_{40}}$ & $\sigma_{\overline{C}_{60}}$ \\
 \noalign{\smallskip}\hline\noalign{\smallskip}
$4\times 10^{-12}$ & $2\times 10^{-12}$ \\
 \noalign{\smallskip}\hline\noalign{\smallskip}

\end{tabular}
\end{table}
\section{The impact of the \virg{imprint} on the LAGEOS-LAGEOS II tests}
A further, crucial step consists of evaluating the impact of such an a-priori \virg{imprint} on the test conducted with the LAGEOS satellites and the combination of \rfr{combi}: if the LAGEOS-LAGEOS II uncancelled combined classical geopotential precession computed with the GRACE-based a-priori \virg{imprinted} even zonals of Table \ref{tibulo2}  is a relevant part of, or it is even larger than the combined Lense-Thirring precession, it will be demonstrated that the doubts concerning the a-priori gravitomagnetic \virg{memory} effect  are founded. It turns out that this is just the case because  \rfr{combi} and Table \ref{tibulo2} yield a
combined geopotential precession whose magnitude is 77.8 mas yr$^{-1}$ ($-82.9$ mas yr$^{-1}$ for $\ell=4$ and $5.1$ mas yr$^{-1}$ for $\ell=6$), i.e. just 1.6 times the Lense-Thirring signal itself. This means that the part of the LAGEOS-LAGEOS II uncancelled classical combined node precessions  which is affected by the \virg{imprinting} by the Lense-Thirring force through the GRACE-based geopotential's spherical harmonics  is as large as the LAGEOS-LAGEOS II combined gravitomagnetic signal itself.

We, now, comment on how Ciufolini and Pavlis  reach a different conclusion. They write in Ref.~\cite{crit2}: \virg{However, the Lense-Thirring effect depends on the third
power of the inverse of the distance from the central body, i.e., $(1/r)^3$, and the
$J_2, J_4, J_6 ...$ effects depend on the powers $(1/r)^{3.5}$, $(1/r)^{5.5}$, $(1/r)^{7.5}$ ... of the
distance; then, since the ratio of the semimajor axes of the GRACE satellites
to the LAGEOS' satellites is $\sim\rp{6780}{12270}\cong
1.8$, any conceivable \virg{Lense-Thirring
imprint} on the spherical harmonics at the GRACE altitude becomes quickly,
with increasing distance, a negligible effect, especially for higher harmonics of
degree $l>4$. Therefore, any conceivable \virg{Lense-Thirring imprint} is negligible
at the LAGEOS' satellites altitude.} From such statements it seems that they compare the classical GRACE precessions to the gravitomagnetic LAGEOS' ones. This is meaningless since, as we have shown, one has, first, to compare the classical and relativistic precessions of GRACE itself, with which the Earth's gravity field is solved for, and, then, compute the impact of the relativistically \virg{imprinted} part of the GRACE-based even zonals on the combined LAGEOS nodes. These two stages have to be kept separate, with the first one which is fundamental; if different satellite(s) Y were to be used to measure the gravitomagnetic field of the Earth, the impact of the Lense-Thirring effect itself on them should be evaluated by using the \virg{imprinted} even zonals evaluated in the first stage.
Finally, in their latest statement Ciufolini and Pavlis write in Ref.~\cite{crit2}: \virg{In addition, in (Ciufolini et al. 1997), it was proved with several
simulations that by far the largest part of this \virg{imprint} effect is absorbed in
the by far largest coefficient $J_2$.} Also such a statement, in the present context, has no validity since the cited work refers to a pre-GRACE era. Moreover, no quantitative details at all were explicitly released concerning the quoted simulations, so that it is not possible to judge by.
\section{Conclusions}\lb{tre}
We have analytically investigated the impact of possible a-priori \virg{imprinting} effects of GTR itself on the ongoing Lense-Thirring tests with the LAGEOS satellites in the gravitational field of the Earth modeled from the dedicated GRACE mission.

The classical part of the terrestrial gravitational potential, acting as a source of major systematic error because of its even zonal harmonic coefficients $\overline{C}_{\ell 0}$, is retrieved from the data of the dedicated satellite-based GRACE mission.  GTR, not explicitly solved for so far in GRACE data analyses, may impact the retrieved even zonals  of the GRACE models at a non-negligible level ($\approx 10^{-10}-10^{-11}$ for $\ell=4,6$), given the present-day level of accuracy ($\approx 10^{-12}$ for $\ell=4,6$). It turns out that the resulting a-priori \virg{imprint} of the Lense-Thirring effect itself on the LAGEOS-LAGEOS II data analysis performed to test it is of the same order of magnitude of the general relativistic signal itself.

Further, more robust tests should rely upon Earth gravity models in which GTR is explicitly solved for.


\begin{thebibliography}{99}
%
\bibitem{Thorne}
K. S. Thorne, ''Gravitomagnetism, Jets in Quasars,
and the Stanford Gyroscope
Experiment'', in  Near Zero: New Frontiers of Physics, J. D. Fairbank, B. S. Deaver,
C. W. F. Everitt and P. F.  Michelson, Eds.  New York: W. H. Freeman and Company, 1988, pp. 573-586.
%
\bibitem{Rin}
W. Rindler, Relativity. Special, General and Cosmological.  Oxford: Oxford University Press, 2001, pp. 195-198.
%
\bibitem{Mash}
B. Mashhoon, ''Gravitoelectromagnetism: A Brief Review'', in The Measurement of Gravitomagnetism: A Challenging Enterprise, L. Iorio, Ed. Hauppauge: Nova, 2007, pp. 29-39.
%
\bibitem{Tar}
M. L. Ruggiero and A. Tartaglia,  Gravitomagnetic effects, Il Nuovo  Cimento B, Vol. 117, No. 7, pp. 743-768, July 2002.
%
\bibitem{Sciaf}
G. Sch\"{a}fer, Gravitomagnetic effects, General Relativity and Gravitation, Vol. 36, No. 10, pp. 2223-2235, October 2004.
%
%\bibitem{ART}
%Einstein A. Die Grundlage der allgemeinen Relativit\"{a}tstheorie. Annalen der Physik 1916;  354: 769-822.
%
%\bibitem{FG}
%Einstein A. Die Feldgleichungen der Gravitation. Sitzungsberichte der K\"{o}niglich Preu{\ss}ischen Akademie der Wissenschaften 1915: 844-7.
%
\bibitem{line}
H. C. Ohanian and R. J. Ruffini, Gravitation and Spacetime. 2nd Edition.  New York: W.W. Norton and Company, 1994, pp. 130-240.
%
%\bibitem{treatise}
%Maxwell JC. A Treatise on Electricity and Magnetism. Clarendon Press: Oxford 1873.
%
\bibitem{PLA}
B. Mashhoon, L. Iorio and H. I. M. Lichtenegger, On the gravitomagnetic clock effect, Physics Letters A, Vol. 292, No. 1-2, pp. 49-57., December 2001.
%
\bibitem{Pfi}
H. Pfister, On the history of the so-called Lense-Thirring effect, General Relativity and Gravitation, Vol. 39, No. 11, pp. 1735-1748,  November 2007.
%
\bibitem{LT}
J. Lense and H. Thirring,  \"{U}ber den Einflu{\ss} der Eigenrotation der Zentralk\"{o}rper auf die Bewegung der Planeten und Monde nach der Einsteinschen Gravitationstheorie, Physikalische Zeitschrift, Vol. 19, pp. 156-163, 1918.
%
\bibitem{Pu}
G. E. Pugh, Proposal for a satellite test of the Coriolis prediction of general relativity WSEG Research Memorandum No. 11. The Pentagon: Washington DC: The Pentagon, November 1959.
%
\bibitem{Sci}
L. I. Schiff, Possible new experimental test of general relativity theory, Physical Review Letters, Vol. 4, No. 5,  pp. 215-217, March 1960.
%
\bibitem{gpb}
C. W. F. Everitt, S. Buchman, D. B. DeBra, G. M. Keiser, J. M. Lockhart, B. Muhlfelder, B. W. Parkinson, J. P. Turneaure, and other members of the Gravity Probe B team, Gravity Probe B: Countdown to launch, in  Gyros, Clocks, Interferometers...: Testing
Relativistic Gravity in Space, C. L\"{a}mmerzahl, C. W. F. Everitt and  F. W. Hehl, Eds. Berlin: Springer, 2001,
pp. 52–82.
%
\bibitem{ANA}
C. W. F. Everitt  M. Adams, W. Bencze, S. Buchman, B. Clarke, J. W. Conklin, D. B. DeBra, M. Dolphin, M. Heifetz, D. Hipkins, T. Holmes, G. M. Keiser, J. Kolodziejczak, J. Li, J. Lipa, J. M. Lockhart, J. C. Mester, B. Muhlfelder, Y. Ohshima, B. W. Parkinson, M. Salomon, A. Silbergleit, V. Solomonik, K. Stahl, M. Taber, J. P. Turneaure, S. Wang and P. W. Worden, Gravity Probe B Data Analysis, Space Science Reviews, doi:10.1007/s11214-009-9524-7, 2009.
%
\bibitem{Ginz0}
V. L. Ginzburg, The use of artificial earth satellites for verifying the general theory of relativity, Uspekhi Fizicheskikh Nauk (Advances in Physical Science), Vol. 63, No. 1, pp. 119-122, 1957.
%
\bibitem{Ginz1}
V. L. Ginzburg, Artificial Satellites and the Theory of Relativity, Scientific American, Vol. 200, No. 5, pp. 149-160, May 1959. % satellite LT perigee
%
\bibitem{Ginz2}
V. L. Ginzburg, Experimental Verifications of the General Theory of Relativity, in Recent Developments in General Relativity. London: Pergamon press, 1962, pp. 57-71.
%
\bibitem{Bogo}
A. F. Bogorodskii, Relativistic Effects in the Motion of an Artificial Earth Satellite, Soviet Astronomy, Vol. 3, No. 5, pp. 857-862, October 1959.
%
\bibitem{Cugu1}
L. Cugusi and E. Proverbio, Relativistic effects on the Motion of the Earth's. Satellites,  Journal of Geodesy, Vol. 51, pp. 249-252, 1977.
%
\bibitem{Cugu2}
L. Cugusi and E. Proverbio, Relativistic Effects on the Motion of Earth's Artificial Satellites, Astronomy and Astrophysics, Vol. 69, pp. 321-325, October 1978.
%
\bibitem{Kau}
W. M. Kaula, Theory of Satellite Geodesy. Waltham: Blaisdell, 1966.
%
\bibitem{Ior03}
L. Iorio,
The impact of the static part of the Earth's
gravity field on some tests of General Relativity with Satellite
Laser Ranging,
Celestial Mechanics and Dynamical Astronomy, Vol. 86, No. 3, pp. 277-294, July 2003.
%
\bibitem{Ciuetal96}
I. Ciufolini, D. M. Lucchesi, F. Vespe, and A. Mandiello, Measurement of dragging of inertial frames and gravitomagnetic
field using laser-ranged satellites, Il Nuovo Cimento A, Vol. 109, No. 5, pp. 575-590, May 1996.
%
\bibitem{Ciu96}
I. Ciufolini, On a new method to measure the gravitomagnetic field using two orbiting satellites, Il Nuovo Cimento
A, Vol. 109, No. 12, pp. 1709-1720, December 1996.
%
\bibitem{Ciu04}
I. Ciufolini and E. C. Pavlis, A confirmation of the general relativistic prediction of the Lense–Thirring effect,
Nature, Vol. 431, No. 7011, pp. 958-960, October 2004.
%
\bibitem{CiuPeron}
I. Ciufolini, E. C. Pavlis, and R. Peron, Determination of frame-dragging using Earth gravity models from CHAMP and GRACE, New Astronomy, Vol. 11, No. 8, pp. 527-550, July 2006.
%
\bibitem{Luc07}
D. M. Lucchesi, The Lense Thirring effect measurement and LAGEOS satellites orbit analysis with the new gravity field model from the CHAMP mission, Advances in Space Research, Vol. 39, No. 2, pp. 324-332, 2007.
%
\bibitem{Riesetal}
J. C. Ries, R. J. Eanes, and M. M.  Watkins, Confirming the frame-dragging effect with satellite laser ranging, in  Proceedings of The
16th International Laser Ranging Workshop. \virg{SLR-The Next Generation}, Pozna\'{n} (PL), 13–17 October 2008, S. Schillak, Ed. Available from: http://cddis.gsfc.nasa.gov/lw16/
%
\bibitem{Iornodi}
L. Iorio, The new Earth gravity models and the measurement of the Lense-Thirring effect, in
The Tenth Marcel Grossmann Meeting On Recent Developments in Theoretical and Experimental General Relativity, Gravitation and Relativistic Field Theories. Proceedings of the MG10 Meeting, Rio de Janeiro, Brazil 20-26 July 2003, M. Novello, S. P. Bergliaffa, and R. J. Ruffini, Eds. Singapore: World Scientific, 2006, pp. 1011-1020.
%
\bibitem{Pav02}
E. C. Pavlis,
Geodetic contributions to gravitational experiments in space,
in  Recent Developments in General Relativity:
Proceedings of the 14th SIGRAV Conference on General Relativity and Gravitational Physics (Genova, IT, 18-22 September 2000), R. Cianci, R. Collina, M. Francaviglia, and P.  Fr\'{e} P., Eds.
Milan: Springer, 2002, pp. 217-233.
%
\bibitem{Ries03a}
J. C. Ries, R. J. Eanes, and B. D.  Tapley,
Lense-Thirring
Precession Determination from Laser Ranging to Artificial
Satellites,
in  Nonlinear
Gravitodynamics. The Lense-Thirring Effect, R. J. Ruffini and C.  Sigismondi, Eds.  Singapore: World Scientific,  2003,
pp. 201-211.
%
\bibitem{Ries03b}
J. C. Ries, R. J.  Eanes, B. D.  Tapley, and G. E. Peterson,
Prospects for an Improved Lense-Thirring Test with SLR and the
GRACE Gravity Mission, in
Proceedings of The 13th International Laser
Ranging Workshop, NASA CP (2003-212248), R. Noomen, S. Klosko, C. Noll, and
M. Pearlman, Eds. Greenbelt: NASA Goddard, 2003. Available from:
http://cddisa.gsfc.nasa.gov/lw13/lw$\_${proceedings}.html$\#$science
%
\bibitem{crit1}
L. Iorio, On the reliability of the so-far performed tests for measuring the Lense-Thirring effect with the LAGEOS satellites, New Astronomy, Vol. 10, No. 8, pp. 603-615, August 2005.
%
\bibitem{crit2}
I. Ciufolini and E. C. Pavlis, On the measurement of the Lense–Thirring effect using the nodes of the LAGEOS satellites, in reply to "On the reliability of the so-far performed tests for measuring the Lense-Thirring effect with the LAGEOS satellites" by L. Iorio, New Astronomy, Vol. 10, No. 8, pp. 636-651, August 2005.
%
\bibitem{crit3}
D. M. Lucchesi, The Impact of the Even Zonal Harmonics Secular Variations on the Lense-Thirring Effect Measurement with the two Lageos Satellites, International Journal of Modern Physics D, Vol. 14, No. 12, pp. 1989-2023, 2005.
%
\bibitem{crit4}
L. Iorio, A Critical Analysis of a Recent Test of the Lense–Thirring Effect with the LAGEOS Satellites, Journal of Geodesy, Vol. 80, No. 3, pp. 128-136, June 2006.
%
\bibitem{crit5}
L. Iorio, An assessment of the measurement of the Lense–Thirring effect in the Earth gravity field, in reply to: ''On the measurement of the Lense–Thirring effect using the nodes of the LAGEOS satellites, in reply to "On the reliability of the sofar performed tests for measuring the Lense-Thirring effect with the LAGEOS satellites" by L. Iorio,'' by I. Ciufolini and E. Pavlis, Planetary and Space Science, Vol. 55, No. 4, pp. 503-511, March 2007.
%
\bibitem{IorioSSRV}
L. Iorio, An Assessment of the Systematic Uncertainty in Present
and Future Tests of the Lense-Thirring Effect
with Satellite Laser Ranging, Space Science Reviews, doi:10.1007/s11214-008-9478-1, 2009.
%
\bibitem{IorioCEJP}
L. Iorio, Conservative evaluation of the uncertainty in the
LAGEOS-LAGEOS II Lense-Thirring test, Central European Journal of Physics, Vol. 8, No. 1, pp. 25-32, February 2010.
%
\bibitem{LucBal06}
D. M. Lucchesi and G. Balmino, The LAGEOS satellites orbital residuals determination and the Lense Thirring effect measurement,
Planetary and Space Science, Vol. 54, No. 6, pp. 581-593, May 2006.
%
\bibitem{Luccc07}
D. M. Lucchesi, The LAGEOS satellites orbital residuals determination and the way to extract gravitational and non-gravitational unmodeled perturbing effects,
Advances in Space Research, Vol. 39, No. 10, pp. 1559-1575, 2007.
%
\bibitem{Nor01}
K. Nordtvedt jr.,  Slr contributions to fundamental physics, Surveys in Geophysics, Vol. 22, No. 5-6, pp. 597-602, September 2001.
%
\bibitem{GRACE}
B. D. Tapley and Ch. Reigber, The GRACE mission: status and future plans,  EOS Transactions AGU 2001; 82: Fall Meeting Supplement G41, C-02.
%
\bibitem{eigengrace02s}
Ch. Reigber, R. Schmidt, F. Flechtner, R.  K\"{o}nig , U.  Meyer,
K.-H. Neumayer, P. Schwintzer, and S. Y. Zhu,
An Earth gravity
field model complete to degree and order 150 from GRACE:
EIGEN-GRACE02S,
Journal of Geodynamics, Vol. 39, No. 1, pp. 1-10, January 2005.
%
\bibitem{ggm03}
B. D. Tapley, J. C. Ries, S. Bettadpur, D. Chambers, M. Cheng, F. Condi, and S. Poole, The GGM03 Mean Earth Gravity Model from GRACE, American Geophysical Union, Fall Meeting 2007, abstract $\#$G42A-03.


\end{thebibliography}
\end{document}